\documentclass[twocolumn,aps,prb,floatfix,showpacs]{revtex4}
\usepackage{graphicx,bm}

\begin{document}

\title{
Novel mechanism for weak magnetization with high Curie temperature observed in H-adsorption on graphene
}

\author{J. G. Che}
\affiliation{Surface Physics Laboratory (National Key Laboratory),
Key Laboratory of Computational Physical Sciences (MOE),
Department of Physics and
Collaborative Innovation Center of Advanced Microstructures, Fudan University,
Shanghai 200433, People's Republic of China}


\begin{abstract}
To elucidate the physics underling magnetism observed in nominally nonmagnetic materials with only 
$sp$-electrons, we built an 
extreme model to simulate H-adsorption (in a straight-line form) on graphene. Our first principles 
calculations for the model produce a ferromagnetic ground state with a magnetic moment of one 
Bohr magneton per H atom and a high Curie temperature. The removal of the 
$p_z$-orbitals from sublattice B of graphene introduces $p_z$-vacancies.
The $p_z$-vacancy-induced states are not created from changes in interatomic 
interactions but are created because of a $p_z$-orbital imbalance between two sublattices (A and B) 
of a conjugated $p_z$-orbital network. Therefore, there are critical requirements for the creation 
of these states (denoted as $p_z^{\rm imbalance}$) to avoid further imbalances and minimize 
the effects on the conjugated $p_z$-orbital network. 
The requirements on the creation of $p_z^{\rm imbalance}$ are as follows: 
1) $p_z^{\rm imbalance}$ consists of $p_z$-orbitals of only the atoms in sublattice A, 
2) the spatial wavefunction of $p_z^{\rm imbalance}$ is antisymmetric, 
and 3) in principle, $p_z^{\rm imbalance}$ extends over the entire crystal without 
decaying, unless other $p_z$-vacancies are crossed. 
Both the origin of spin polarization and the 
magnetic ordering of the model arise from the aforementioned requirements.
\end{abstract}

\maketitle
\section{Introduction}
Magnetism in nominally nonmagnetic materials that contain only $sp$-electrons is currently a 
popular research topic because of the potential practical importance of these 
materials\cite{Esq13,Han14,Khe14,Gon19}. Of all 
controversial experimental observations that have been reported for these 
materials\cite{Mak06,Sep10,Nai12,Sep12}, the 
high Curie temperature under such weak magnetization (three or four orders of magnitude smaller 
than that for conventional magnets) is quite confusing\cite{Elf02,Esq03,Cer09,Uge10,Gie13,Wan09,Gon16}. 
There is almost no means of explaining how defect-induced magnetic moments (MMs) that 
are localized far away from each other (where the MMs' distance are related to the defect 
concentration) could be coupled in ferromagnetic (FM) ordering above room temperature. 
Although calculation results seem to 
consistently show that the $sp$-electrons are spin-polarized by 
defects\cite{Esq13,Yaz07,Bou08,San12,Yaz10,Vol10}, no convincing description 
of the physics underlying this phenomenon has been put forward as yet. 
Exploring the origin of this phenomenon has thus become an overwhelming challenge in materials 
science\cite{Gon19,Fis15}.

Two serious difficulties arise in the theoretical study of these materials. First, the origin of 
$sp$-electron spin polarization remains unexplained because Heisenberg established 90 years ago 
that the principle quantum number of electrons that contribute to magnetism must be greater than or 
equal to three\cite{Hei26,Hei28}. 
Second and more critical is that no current theories have convincingly explained 
how long-range localized MMs (on defect centers) can be coupled in FM ordering at such a high 
Curie temperature. 
There must be an unrecognized magnetic mechanism at work.

In our previous papers\cite{Xu19a,Xu19b}, 
we showed that the spatial wavefunctions of the electronic states that 
contribute to magnetism are antisymmetric, thereby dispelling the uncertainties regarding 
$sp$-electron spin polarization. To address the question of magnetic coupling, we proposed in our 
previous paper on magnetism in graphene with vacancies\cite{Xu19a} 
that the $p_z$-orbital imbalance that is 
induced by vacancy would result in magnetic ordering in the material. However, it is 
computationally intensive to simulate graphene with a vacancy concentration that is sufficiently 
similar to that for which magnetism can be experimentally observed\cite{Uge10}. Hence, we did not perform 
first principles calculations to compare the total energy difference between the FM and 
antiferromagnetic (AFM) ordering but only performed an analysis that led to the abovementioned 
conclusions.

In a two-dimensional (2D) system, a point defect leads to a resonance that decays with $r$ (the 
distance from the defect)\cite{Xu19a,Per06,San10,Pal12,Sun17}, 
whereas according to perturbation theory\cite{Pol80}, a defect that is periodically 
arranged infinitely throughout the entire 2D system can create a resonance without a decay. In this 
paper, we build an extreme model with H-adsorption (in a straight-line form) on graphene to 
demonstrate robustly that the FM (with a high Curie temperature) originates from a 
$p_z$-orbital imbalance between two sublattices of graphene. Our aim is not to determine whether 
such an extreme model could be experimentally realized; rather, we wish to use the extreme model 
to better understand the novel magnetic mechanism\cite{Xu19a} that is at work in these materials.

\section{Calculation methods}
Our results were obtained using first principles calculations that were implemented using the VASP 
package\cite{vasp} with the same calculation setup as in our previous papers\cite{Xu19a,Xu19b}, 
that is, the wavefunctions were 
expanded in a plane-wave basis set with an energy cutoff of 500~eV. The interaction between the 
atoms and electrons was described by the projector augmented plane-wave method\cite{PAW}. The 
generalized gradient approximation\cite{GGA-PBE} for exchange-correlation effects was used. 
The H-adsorption on 
graphene was simulated using a supercell in a 48$\times$1-sized cell of the original graphene. The 
k-points were sampled in the 2D Brillouin zone on 4$\times$48 meshes for the 
total energy calculations. Vacuum thickness was 
maintained larger than 20~\AA. All of the atoms were relaxed until the Hellmann-Feynman forces 
on the atoms were smaller than 0.02~eV/\AA. This calculation setup was found to be sufficiently 
accurate for the purposes of our study. For example, the use of this calculation setup resulted in a 
calculated lattice constant for graphene of 1.42~\AA, which is in good agreement with the 
experimental value\cite{Net09}.

\section{Results and discussion}
\subsection{one H adsorption on graphene}

\begin{figure}[bt]
\centerline{\includegraphics[scale=0.18,angle=0]{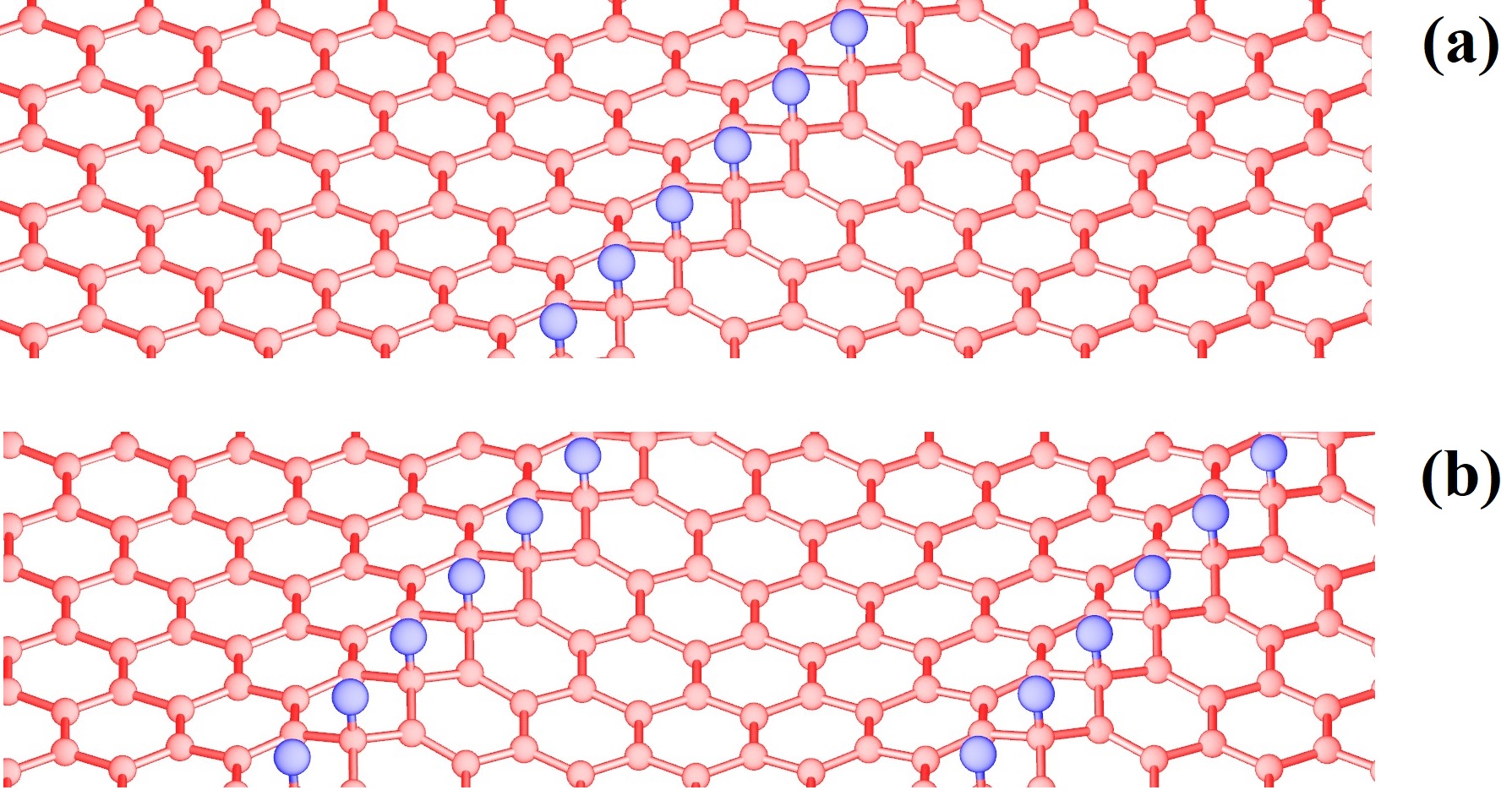}}
\caption{(Color online)
Perspective view of H-adsorption on one sublattice of graphene forming (a) one and (b) two 
H lines. Blue and red balls represent H and C atoms, respectively. In (b), two H lines are separated 
by six of the original graphene units
}
\label{config}
\end{figure}

Bonded crystals with only $sp$-electrons do not exhibit magnetism, as concluded by Heisenberg 
theory\cite{Hei26,Hei28}. 
However, magnetism that is induced by vacancies or nonmetal adatoms (H, F, etc.) on 
graphene and graphite has been experimentally observed above room temperature and predicted by 
calculations\cite{Esq13,Kuz13,Han14,Fis15}. 
However, a convincing theory that 
explains these phenomena is still required\cite{Han14,Yaz10,Fen17,Naf17,Kuz13,Fis15,Kat12,Sin13}. 
Previously, we have explained\cite{Xu19a,Xu19b} why Hund's rule does not 
hold when interpreting calculation results based on the singlet-electron approximation and the 
Bloch theorem in band theory.

In a previous study\cite{Xu19a}, we proposed that the $p_z^*$ -nonbonding states (which is called the 
zero-mode in the literature\cite{Per06,Per08}) play a critical role in both $sp$-electron 
spin polarization and magnetic 
coupling in graphene with vacancies. As mentioned in the Introduction, in the present study, an 
extreme model was built to simulate H-adsorption (in a straight-line form) on graphene, as shown 
in Fig. 1 (a). Instead, of introducing a C-vacancy by removing a C atom in graphene, we introduced 
a $p_z$-vacancy in graphene via the H-saturation of a $p_z$-orbital of C below H (denoted as 
C$_{\rm H}$). This model enabled us to avoid cutting off graphene (if the vacancies form in a straight line) 
while excluding antibonding states (that are induced by the interactions among the three dangling 
bonds that were left by the vacancy). The model enabled us to focus only on the 
$p_z$-vacancy-induced states. Considering our computational resources, we adopted a supercell of 
a 48$\times$1-sized graphene to observe the long-range nature of the resonance states and examine 
the coupling of MMs in the model.

The C($2s^22p^2$) atom in graphene prefers to hybridize in one $p_z$ and three $sp^2$ orbitals 
(where the $z$-axis is perpendicular to the graphene plane)\cite{Net09}. 
The $sp^2$ orbitals form $\sigma$-bonds as a 
backbone of the honeycomb lattice that is composed of two sublattices that are denoted by A and B. 
Each of the atoms in sublattice A has three neighboring atoms in sublattice B and vice versa. The $p_z$ 
and three $sp^2$ orbitals in graphene are orthogonal and thus do not interact with each other. 
Therefore, the $p_z$-orbitals ($p_z$-electrons) over the two sublattices under C$_{\rm 3V}$ symmetry form a 
conjugated $p_z$-orbital ($p_z$-electrons) network in a coherent manner, 
which is the critical factor for producing magnetism in graphene.

Fig. 1 (a) shows the atomic configuration of the supercell for the adsorption of one H atom onto the 
48$\times$1-sized graphene with unit vectors, ${\bf a}_1=48(\sqrt{3/2}, -1/2)a$ 
and ${\bf a}_1=48(\sqrt{3/2}, 1/2)a$. 
C$_{\rm H}$ has three carbon atoms as its nearest neighbors. However, only two of the three carbon atoms 
could lie in the supercell because of the period along ${\bf a}_2$. The optimized atomic structure shows that 
the adsorption of H caused C$_{\rm H}$ to lift 0.68~\AA\ above the graphene plane, whereas the two 
nearest neighbors of the H atom were at 0.34 and 0.26~\AA\ above the graphene plane. The 
distances between C$_{\rm H}$ and its two nearest neighbors were 1.48 and 1.52~\AA. For comparison, the 
corresponding value in a perfect graphene is 1.42~\AA\cite{Net09}. Clearly, C$_{\rm H}$ was transformed from $sp^2$ 
hybridization in graphene to $sp^3$ hybridization after H-adsorption, and its $p_z$-orbital 
(dangling bond) was saturated as expected, introducing to a $p_z$-vacancy, that is, there was a 
$p_z$-orbital imbalance between the two sublattices. Except for the aforementioned changes, there 
were no other changes in the C-C bond lengths that were larger or smaller than 0.04~\AA\ relative 
to those in a perfect graphene.

\begin{figure*}[bt]
\centerline{\includegraphics[scale=0.34,angle=0]{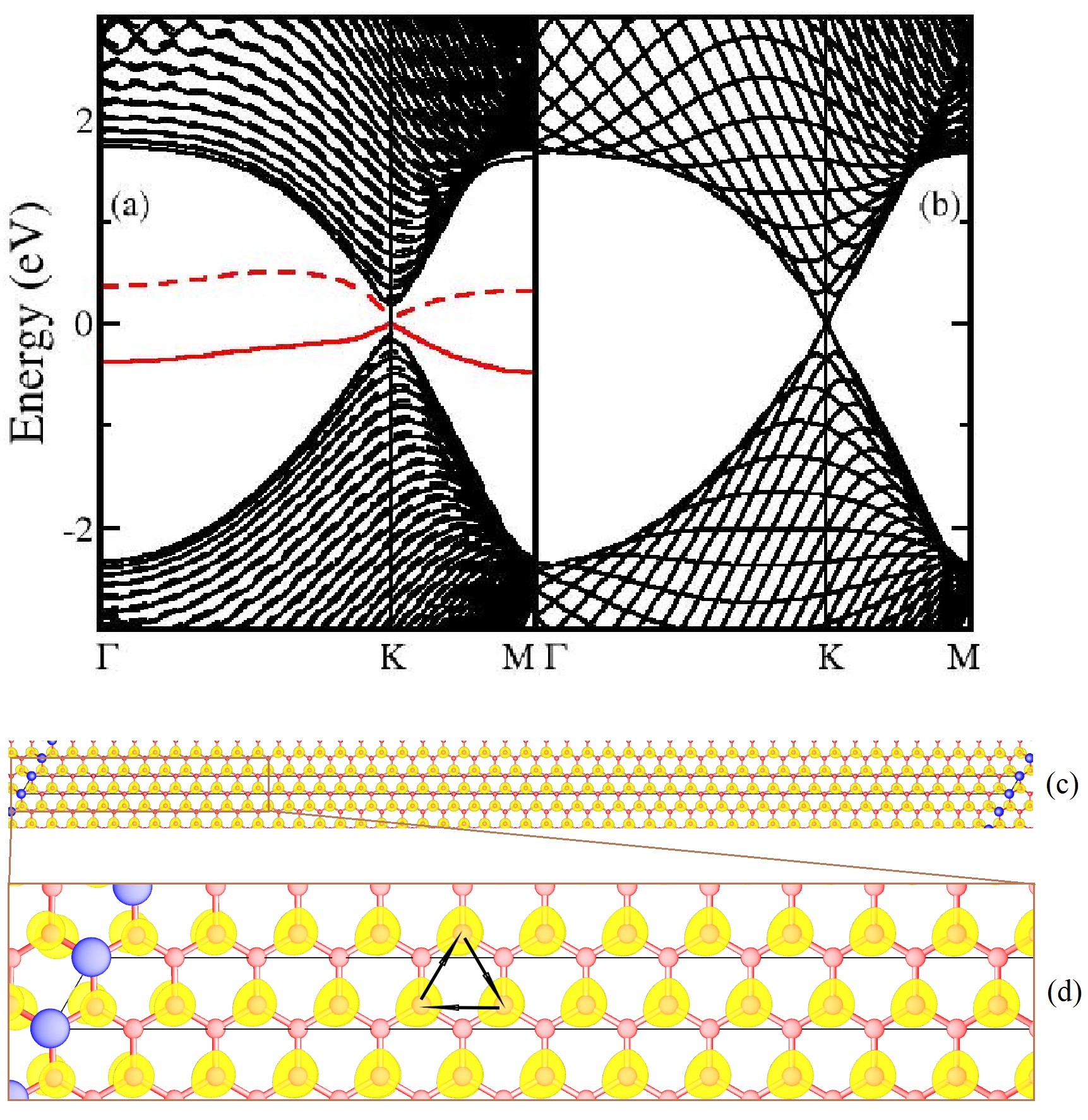}}
\caption{(Color online)
Band structures of a 48$\times$1-sized graphene (a) with and (b) without H-adsorption, 
where the solid and dashed curves represent majority and minority states, respectively. K (0, 1/3) 
with ${\bf b}_1$ and ${\bf b}_2$ as units is folded from the K-point of the original graphene, 
and the Fermi level is 
set to zero. (c) Charge distribution of the solid red band in (a) at the K-point. (d) Enlargement of 
the brown rectangular portion in (c). Thick lines in (c) and (d) represent the 2D boundary of the 
supercell, red and blue balls in (c) and (d) represent C and H atoms, respectively, and the yellow 
regions show the charge distribution for the red state at the K-point in (a). Arrows in (d) show the 
phase difference between the two respective atoms.
}
\label{band-1H}
\end{figure*}

To determine the origin of the magnetism, the band structures for the ground state of the 
48$\times$1-sized graphene with and without H-adsorption are shown in Figs. 2 (a) and (b), 
respectively. Comparing the two band structures, there are two features after H-adsorption on the 
48$\times$1-sized graphene. First, the degeneration at the Dirac-point of graphene, as shown in Fig. 
2 (b), was lifted after H-adsorption, creating a gap of approximately 0.3~eV. Second, the red (solid 
and dashed) bands in Fig. 2 (a) exhibit spin-splitting from 0.1 to 0.9~eV along 2D-BZ. The graphene 
with a $p_z$-vacancy was thus semiconducting with a gap of 0.1~eV. The majority (solid) red band 
was fully occupied, and the minority (dashed) red band was empty, contributing 1~$\mu_{\rm B}$ 
per H to the MM. Fig. 2 (b) clearly shows the folded structure of the original graphene band that 
resulted from a 48-folded 2D-BZ. 

In previous studies\cite{Yaz07,Bou08,San12,Pal12,Gon16}, 
the H atom was considered to be a point-defect 
adsorption on graphene, and the induced 1~$\mu_{\rm B}$ per H was ascribed to Hund's rule 
because the 
induced electron states appeared to be quite localized. However, as discussed in our previous 
papers\cite{Xu19a,Xu19b}, 
caution is required when using first principles calculations that are based on band theory 
because all of the electrons in band theory extend over the entire crystal as per the Bloch theorem, 
even if the bands appear to be flat\cite{Bloch}.


To examine the nature of the majority red band, we calculated the charge distribution of the 
majority red band at the K-point, which is shown in Figs. 2 (c) and (d). At first glance, the electrons 
were distributed on each of the atoms in sublattice A with almost the same 
isosurfaces, indicating that the red state extended over the entire supercell without decaying, as 
expected, which motivated us to study straight-line defects in graphene instead of point defects. In 
fact, the electron distribution on only sublattice A indicated an important feature: the spatial 
wavefunction of the state should be antisymmetric. That is, the phases of the wavefunction components 
on the two 
neighboring atoms belonging to sublattice A should be antisymmetric. Otherwise, electrons would 
accumulate on the atom (sublattice B) between the two neighboring atoms (sublattice A),
resulting in an extra imbalance. Therefore, it is impossible. Note that 
the length of the 48$\times$1-sized supercell was already at 120~\AA. In principle, a consequence 
of the antisymmetric wavefunction is that the state should extend infinitely. Otherwise, terminating 
the wavefunction at one point will create an additional imbalance without any wavefunction at the 
opposite side of the broken point.

Although the red state in Fig. 2 (a) was an H-induced state, its wavefunction was not localized but 
extended over the entire supercell and consisted of $p_z$-orbitals of only the atoms in sublattice A. 
Note that the atoms in sublattice A belonged to the second nearest neighbors\cite{Net09}. 
Therefore, the state 
was not caused by a change in the localized atomic interaction due to H-adsorption, but was created 
as a response of the conjugated $p_z$-electron network to a $p_z$-electron imbalance between the 
two sublattices.

In our previous paper\cite{Xu19a}, we used "nonbonding state ($p_z^*$)" to describe the involved state 
following molecular 
orbital theory from quantum chemistry. However, imbalance is more essential than nonbonding in 
the inherent nature of this state. In the absence of a suitable name, we temporarily refer to this state
as an imbalanced state (denoted as $p_z^{\rm imbalance}$) for the purposes of discussion in the following section, 
to distinguish this state from the lone pair dangling bond state, which
is also referred to as a nonbonding state in the literature.

As the induced state was caused by a $p_z$-electron imbalance, 
the response of the conjugated $p_z$-electron network to the imbalance 
should not create any further imbalance and should minimize the effect on the conjugated 
$p_z$-electron network. When one $p_z$-electron in sublattice A loses its pairing in sublattice B, 
the imbalance could be rectified by filling the induced state with one $p_z$-electron in sublattice A. 
However, if one $p_z$-electron was provided by one C atom in sublattice A (denoted by C$_{\rm A}$) near 
the $p_z$-vacancy, this response would be both electrostatically unstable and break its pairing with 
the atom in sublattice B near C$_{\rm A}$. Thus, unpairing would continue indefinitely, alternating on the 
two sublattices. Therefore, to minimize the effect of the response on the conjugated $p_z$-electron 
network, each atom in sublattice A approximately provided $1/N_{\rm A}$ electrons to form the 
$p_z^{\rm imbalance}$-imbalanced state, where $N_{\rm A}$ is the number of atoms in sublattice A in the supercell.

Therefore, rectifying the imbalance involved the following requisites: 1) $p_z^{\rm imbalance}$ consisted of 
$p_z$-orbitals of only the atoms in sublattice A, 2) the $p_z^{\rm imbalance}$ wavefunction was antisymmetric, 
and 3) $p_z^{\rm imbalance}$ extended infinitely over the entire crystal without decaying. The electron 
antisymmetric exchange principle\cite{QM} states that the spin wavefunction of $p_z^{\rm imbalance}$ should be 
symmetric.  This is the origin of the MMs caused by the $sp$-electrons
in this material. That is, the MMs 
originated from the response of the conjugated $p_z$-electron network to the $p_z$-electron 
imbalance between two sublattices.

Counting electrons within the supercell, only one electron filled $p_z^{\rm imbalance}$ and was distributed over 
all of the 48 atoms of sublattice A of the entire supercell, which corresponded to a contribution of 
1~$\mu_{\rm B}$. From the isosurfaces that are shown in Figs. 2 (b) and (c), we concluded that the 
charge distribution on these atoms was approximately the same. 
That is, there was a contribution of 1/48~$\mu_{\rm B}$ from each atom of sublattice 
A in the supercell. Note that these MMs were not atomic MMs isolated on atoms but were 
inherently parallel on the atoms because the MMs were the collective contribution from one 
electronic state. This result explains why such weak MMs with only 1~$\mu_{\rm B}$ and in 
parallel alignment could be distributed so widely over the entire supercell.

Here, we should explain the implications of an antisymmetric wavefunction for the supercell, as is shown 
in Fig. 1 (a). There was a phase difference of $e^{-i2\pi/3}$ between neighboring atoms in 
sublattice A. The angle between ${\bf a}_2$ and ${\bf a}_1$ was $2\pi/3$. Each arrow in Fig. 2 (d) represents a phase 
factor of $e^{-i2\pi/3}$ between the two respective atoms. 
Thus, an antisymmetric wavefunction means that the sum of the $p_z$-orbital components of the 
three atoms surrounding any atom in sublattice B was zero (within an accuracy of $10^{-6}$), thereby 
guaranteeing that there were no orbital components for $p_z^{\rm imbalance}$ on sublattice B.

\subsection{Two H adsorption on graphene}

A single $p_z$-vacancy in sublattice B created $p_z^{\rm imbalance}$, which was the response of the conjugated 
$p_z$-electron network to a $p_z$-electron imbalance between two sublattices. 
How did the conjugated $p_z$-electron network respond to two 
$p_z$-vacancies simultaneously on sublattice B? 
More $p_z$-electron in 
sublattice A were required to rectify the imbalance caused by increasing the $p_z$-vacancies in 
sublattice B because $p_z^{\rm imbalance}$ was not created by a change in localized atomic interactions but from 
a $p_z$-electron imbalance. 
The requirements (no additional imbalance and minimized the effect on the conjugated $p_z$-electron network) 
on $p_z^{\rm imbalance}$ resulted in only a recombination of the $p_z$-orbitals of sublattice A 
for this case.

\begin{table}[h]
\caption{
Total energy (eV) per supercell for FM and AFM of two H separated by units ($N$).
$\Delta E=E({\rm AFM})-E({\rm FM})$
}
\label{energy}
\begin{ruledtabular}
\begin{tabular}{lccccc}
$N$       & 3      & 6      & 12     & 18     & 24 \\ \hline
$E$(FM)   &-889.365&-889.427&-889.419&-889.422&-889.427\\
$E$(AFM)  &-889.338&-889.404&-889.398&-889.400&-889.405\\
$\Delta E$&0.027   &0.022   &0.022   &0.022   &0.022
\end{tabular}
\end{ruledtabular}
\end{table}

The conclusion was obtained by considering two $p_z$-vacancies separated by different units ($N$) 
on sublattice B.  The atomic 
configuration of two H atoms on one sublattice ($N$ = 6) is shown in Fig. 1 (b). 
The configurations for $N$ = 3, 6, 12, 18, and 24 were similar. Table I lists the total energies of FM 
and AFM for different $N$. From the table, FM was energetically more favorable than AFM by at 
least 220 meV. 
Because the current models calculating the transition temperature are based on the magnetic coupling 
(interaction parameters) between the nearest neighbors, we cannot use them for our cases, 
in which one MM contributed of one electronic state widely distributes as a whole moment 
on all atoms of sublattice A (see below). 
However, we would point out that the exchange energy of 220meV corresponds 
(1~eV = 1.16048*10$^4$~K)
to a temperature higher than 2500K, in agreement with experimental observations 
of a high Curie temperature in proton-irradiated graphene. 
For all of the five cases, the MM for FM was 2~$\mu_{\rm B}$ per 
supercell, or 1~$\mu_{\rm B}$ per H.

\begin{figure*}[tb]
\centerline{\includegraphics[scale=0.90,angle=0]{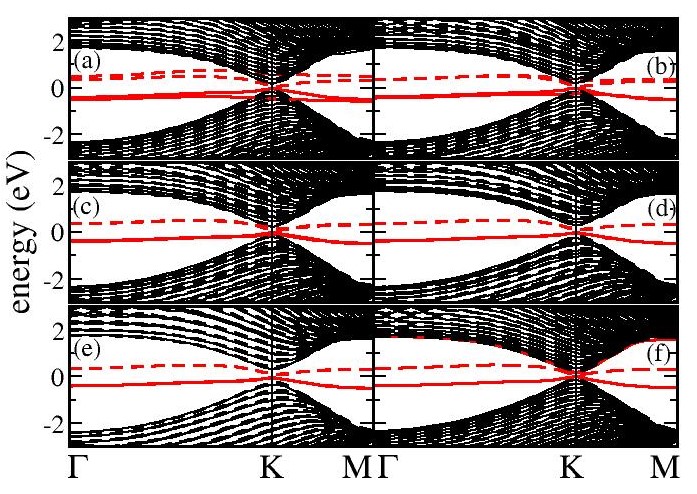}}
\caption{(Color online)
Same caption as that for Fig. 2 (a), except for band structures of the 48$\times$1-sized 
graphene with two adsorbed H that are separated by (a) 3, (b) 6, (c) 12, (d) 18, and (e) 24 units; (f) 
is the same as Fig. 2 (a) and serves as a comparison.
}
\label{band-2H}
\end{figure*}

The band structures of two H atoms separated by $N$ = 3, 6, 12, 18, and 24 units adsorbed on one 
sublattice of the 48$\times$1-sized graphene are plotted in Figs. 3 (a)-(e), respectively. The band 
structure of one H-adsorbed on the same model is also plotted in Fig. 3 (f) for comparison. First, it 
can be clearly seen that the MMs of 2~$\mu_{\rm B}$ per supercell 
for these systems came from the two red bands in Figs. 
3 (a)-(e). Second, both the dispersion and energy level of these 
red bands in Figs. 3 (a)-(e) were similar to that of Fig. 3 (f), except for the small splitting of the two 
red bands for $N$ = 3 and $N$ = 6 near the K-point. The splitting was attributed to the charge 
distributions for the two red bands and is discussed below.

\begin{figure*}[bt]
\centerline{\includegraphics[scale=0.36,angle=0]{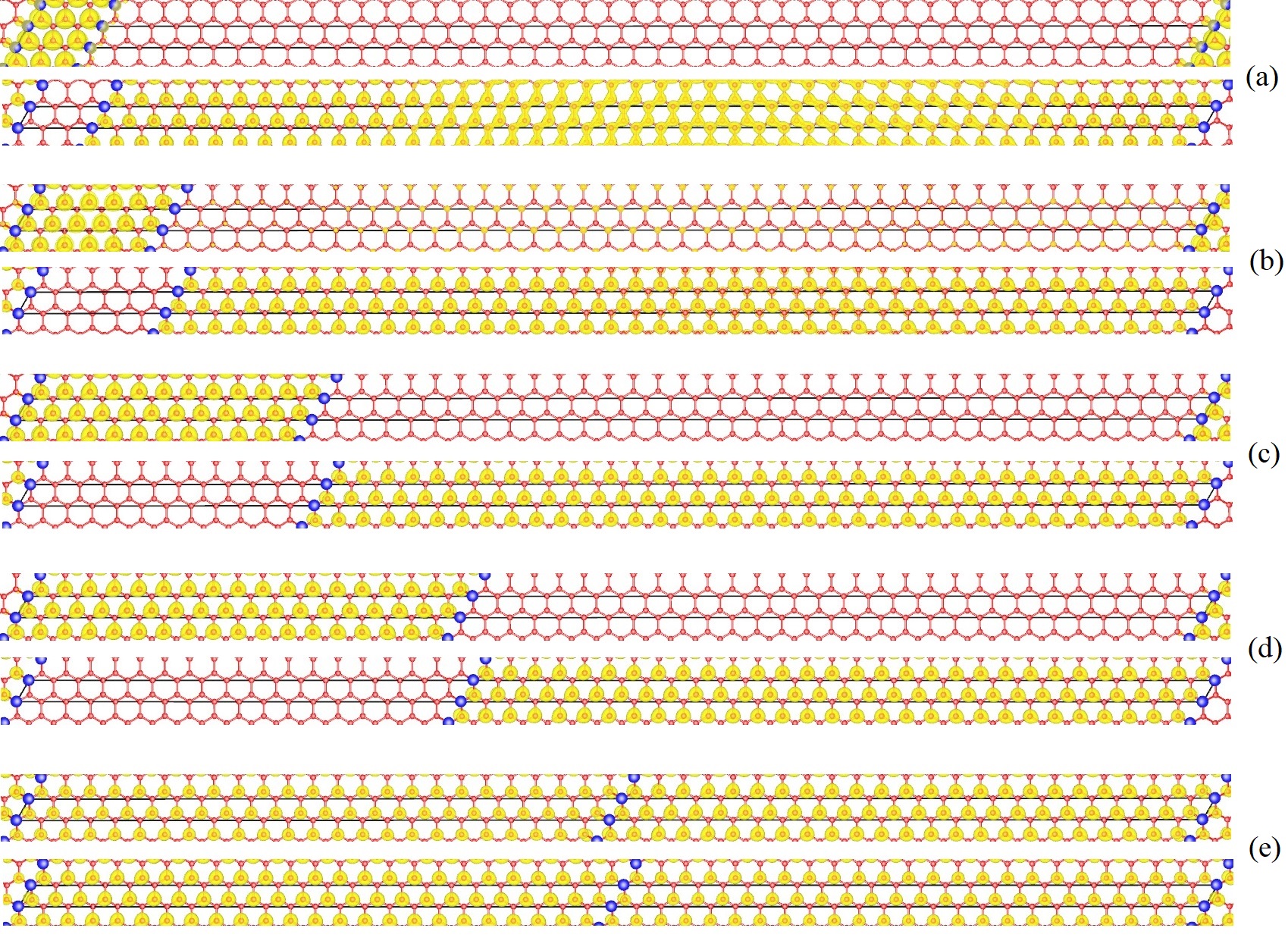}}
\caption{(Color online)
Charge distribution of two occupied states that correspond to the red bands at the 
K-point in Figs. 3 (a)-(e).
One H atom is adsorbed on unit 0 of the 48$\times$1-sized graphene, 
and (a), (b), (c), (d), and (e) correspond to the other H atom 
adsorbed on units 3, 6, 12, 18, and 24, respectively. Top and bottom subfigures in each panel 
correspond to imbalanced states with lower and higher energies, respectively. Red and blue balls 
represent C and H atoms, respectively, and thin solid lines represent the supercell boundary.
}
\label{charge-2H}
\end{figure*}

The charge distributions for the two solid red bands in Figs. 3 (a)-(e) at the K-point are shown in 
Figs. 4 (a)-(e), respectively. 
The top and bottom subfigures in each panel of Figs. 4 (a)-(e) correspond to 
two imbalanced states: the top subfigure had a lower energy than the bottom subfigure. 
From Figs. 4 (a)-(d), it was found that the 48$\times$1-sized supercell could be divided into two segments 
with the H atom ($p_z$-vacancy) as the boundary: 
short and long segments with lengths of $N$ (3, 6, 12, and 18) and $48-N$ units, respectively. 
The electrons that occupied each of the red bands were distributed on 
only one of the two segments, leaving the other segment empty, 
and the electrons were evenly distributed on each atom of sublattice A within the segment. 
These results showed that the requirement for the antisymmetry of the wavefunction was fulfilled. 

The splitting of the two red bands that 
appeared near the K-point for the cases $N$ = 3 and $N$ = 6 was caused by the difference in the 
exchange energy between the short and long segments, because
the exchange energy depended on the charge density of the involved atoms\cite{magnetism}.
One electron was evenly distributed on each atom of sublattice A within the segment, 
$p_z^{\rm imbalance}$ distributed in the short segment had thus a lower energy than that in the long segment. 
The band-splitting for the cases $N$ = 12, 18, and 24 was smaller than 0.02 eV.

It is shown in Figs. 4 (a)-(d) that the $p_z$-vacancy acted as a boundary in dividing the supercell 
into two segments, that is, the charge distribution for $p_z^{\rm imbalance}$ did not have the $p_z$-vacancy as its 
center, although $p_z^{\rm imbalance}$ was induced by the $p_z$-vacancy. This also implies that $p_z^{\rm imbalance}$ 
resulted from an imbalance, not an interaction. Most importantly, this characteristic led to the 
recombination of the $p_z$-orbitals in sublattice A, thereby forming individual segments for each 
of the $p_z^{\rm imbalance}$ states to rectify the imbalance.

However, for $N$ = 24, the electron distribution in the supercell could not be divided into two 
segments. Instead, the electrons of both red states were distributed over the entire supercell, as 
shown in the top and bottom subfigures of Fig. 4 (e). The two red bands were almost degenerate, as in 
the case of the single H atom that was adsorbed on the 24$\times$1-sized graphene. We performed
calculations for the case $N$ = 24 with two different initial atomic configurations:
1) two H atoms were placed on the original graphene, and 2) one H atom was 
placed on the relaxed graphene with one adsorbed H atom. 
The two initial configurations produced the 
same ground state: the total energy and MMs were the same at the level of accuracy of the 
calculation, and the DOS and charge distribution were similar. Most importantly, the electrons of 
the two red bands beginning with both initial configurations were distributed over the entire supercell, 
unlike for the cases of $N$ = 3, 6, 12, and 18, for which the electrons were distributed over one 
segment of the supercell, leaving the other segment empty.

The analysis above led us to conclude that the response of the conjugated $p_z$-electron network 
to two $p_z$-vacancies on one sublattice was similar to that for one $p_z$-vacancy because the 
origin of $p_z^{\rm imbalance}$ was the same for one or two $p_z$-vacancies. 
Regardless of the number of $p_z$-vacancies that 
appear on one sublattice, the aforementioned requirements should always be fulfilled. Therefore, 
constructing the wavefunction for each of the two $p_z^{\rm imbalance}$ states corresponded simply to a 
recombination of the $p_z$-orbitals in sublattice A.

This conclusion is borne out by Figs. 4 (a)-(d): two red states ($p_z^{\rm imbalance}$), respectively, are 
distributed on two individual segments without overlapping with each other, as shown in the top 
and bottom subfigures in each of Figs. 4 (a)-(d). The two individual segments shared a common 
$p_z$-vacancy as a boundary. 

The magnetic ordering of the MMs in the two segments is illustrated in Fig. 5. In 
the figure, the two segments are differentiated by color and black-white lobes. The up-down colors 
(black-gray) of the alternative lobes indicate antisymmetric wavefunctions. 
In principle, the wavefunction phases of the two $p_z^{\rm imbalance}$ states should be independent, 
that is, the up-down colors of the lobes in the
left segment could be inverted relative to the black-gray up-down order in the right segment, as 
shown in Figs. 5 (a) and (b).

Let up-pink and down-green (up-black and down-gray) denote the positive phase: then, if the 
orbital phases on two atoms on the two sides of the $p_z$-vacancy (denoted by C$_{\rm left}$ and 
C$_{\rm right}$ respectively) were antisymmetric, 
as shown in Fig. 5 (a), no electrons accumulated on the $p_z$-vacancy site. 
Otherwise, electrons could have appeared on the $p_z$-vacancy-site, 
if the orbital phases on  C$_{\rm left}$ and C$_{\rm right}$ are symmetric, as shown in Fig. 5 (b). 
Therefore, the condition that no additional imbalance was allowed for $p_z^{\rm imbalance}$ required the 
orbital phases on C$_{\rm left}$ and C$_{\rm right}$ to be antisymmetric, thereby favoring 
parallel alignment between the two MMs (blue arrows) on C$_{\rm left}$ and C$_{\rm right}$.
The white arrows on the other atoms in each of two segments followed the individual blue arrows, 
leading to all the MMs in the two segments being aligned parallel to each other because the MMs 
(blue and white arrows) within the same segment belonged to one electronic state, 
which should be parallel. 
By contrast, Fig. 5 (b) shows that the two boundary atoms,
C$_{\rm left}$ and C$_{\rm right}$, had symmetric orbital
phases; thus, the MMs on the two boundary atoms (blue arrows) should have been antiparallel. For the 
case $N$ = 24, the MM alignment could be easily understood: the charge distributions of the two 
$p_z^{\rm imbalance}$ states overlapped; thus, the MMs of one $p_z^{\rm imbalance}$ state could act as an MM field to induce the 
MMs of the other $p_z^{\rm imbalance}$ state to align in parallel.

\begin{figure}[bt]
\centerline{\includegraphics[scale=0.40,angle=0]{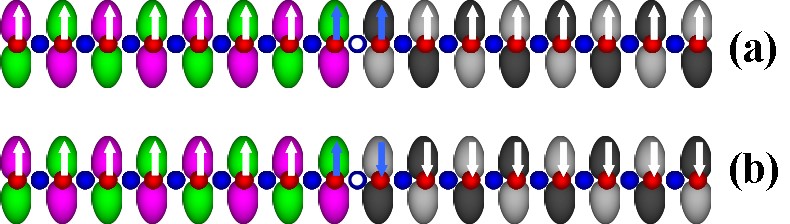}}
\caption{(Color online)
Schematic of the relationship between the phase and magnetic ordering between two 
$p_z^{\rm imbalance}$-dominant segments: red and blue balls represent atoms of sublattices A and B, respectively; 
the blue ball with a white circle corresponds to the position of the $p_z$-vacancy; colored and 
black-white lobes belong to two $p_z^{\rm imbalance}$-dominant segments that share a common $p_z$-vacancy 
as a boundary; colored lobes with up-pink and down-green, and black-white lobes with up-black 
and down-gray indicate positive phases; blue arrows indicate MMs on two atoms of two sides of
the $p_z$-vacancy; and white arrows are MMs on the other atoms of segments.
}
\label{coupling}
\end{figure}

Considering that the $p_z^{\rm imbalance}$'s electron was
evenly distributed on the atoms of sublattice A 
and there were $N_{\rm AL}$ and $N_{\rm AS}$ atoms of sublattice A 
within the long and short segments respectively, 
the two boundary atoms, C$_{\rm left}$ and C$_{\rm right}$, 
had thus $1/N_{\rm AL}$ and $1/N_{\rm AS}$ electrons. 
According to the above analysis about the orbital phases on the two boundary atoms, 
the coupling of the MMs in the two segments depended on 
the $1/N_{\rm AL}$ and $1/N_{\rm AS}$ electrons.
The lower the $p_z$-vacancy concentration, the smaller the $1/N_{\rm AL}$ 
($1/N_{\rm AS}$) electrons, and the lower the exchange energy. 
Compared with the case of a point defect, such as graphene with 
vacancies, the model presented here represents an extreme case in which a state without decay can 
be realized. 
In the point-defect case, such as a vacancy in graphene, the charge density of a defect 
state can fast decay to a value that is smaller than a thermal fluctuation on atoms of sublattice B. 
This difference between point-defects and straight-line-defects may explain the occurrence of 
controversial experimental observations\cite{Mak06,Sep10,Nai12,Sep12}.

\section{Conclusions}
In summary, we studied magnetism that has been observed in proton-irradiated graphene by 
constructing an extreme model of a supercell for an H-adsorbed (modeled as a straight line) onto 
graphene. We present a novel mechanism to explain the magnetic phenomena.
This mechanism was first suggested in our previous papers\cite{Xu19a}, 
and it is substantially different from conventional models such as the Heisenberg model, the 
indirect exchange model, the superexchange model, the RKKY model, and the itinerant electron 
model\cite{magnetism}.

We showed that the conjugated $p_z$-electron network in graphene plays a critical role in 
$sp$-electron spin polarization and magnetic ordering. The H-adsorption on graphene saturates a 
$p_z$-orbital of C under the H atom, creating a $p_z$-vacancy on the conjugated $p_z$-electron 
network, thereby inducing a $p_z$-electron imbalance between two sublattices. As the $p_z^{\rm imbalance}$ 
state is caused by an imbalance, no further imbalance and minimized effect on 
the conjugated $p_z$-electron network should be required for the creation of $p_z^{\rm imbalance}$. 
That is, 
1) the state consists of $p_z$-orbitals only on the atoms in sublattice A, 
2) the spatial wavefunction of the state is antisymmetric, and 
3) the state extends over the entire crystal without decay unless other $p_z$-vacancies are crossed. 
The $sp$-electron spin polarization in graphene with a $p_z$-vacancy originates from the spatial 
antisymmetric wavefunction of $p_z^{\rm imbalance}$ as per the electron exchange antisymmetric principle.

An increase in $p_z$-vacancies results in the creation of more $p_z^{\rm imbalance}$ states because $p_z^{\rm imbalance}$ 
states are created from a $p_z$-orbital imbalance and not from interatomic interactions. Although 
the response of the conjugated $p_z$-electron network to the imbalance is to create more 
$p_z^{\rm imbalance}$ states, the requirement of the formation of the induced states must still be fulfilled, 
resulting in only the recombination of $p_z$-orbitals in sublattice A. 
If two $p_z$-vacancies exist, 
the electrons that fill up each of two $p_z^{\rm imbalance}$ states are not distributed over the entire crystal but 
in two individual segments where the $p_z$-vacancies serve as boundaries. 
The wavefunction phase of 
each state should be independent, although all of the $p_z^{\rm imbalance}$ states consist of $p_z$-orbitals of 
sublattice A in each segment that are antisymmetric. 
However, the requirements that no additional imbalance should be introduced and that the effect 
on the network should be minimized result in the orbital phases on two boundary atoms,
C$_{\rm left}$ and C$_{\rm right}$, to be 
antisymmetric, leading to the presence of MMs on the two boundary atoms in FM alignment with a high
Curie temperature. Therefore, both the spin 
polarization and magnetic ordering that is produced by our extreme model originate from the 
requirements for the imbalanced states.

Because $p_z^{\rm imbalance}$ is filled by only one electron but has a long-range extent and is distributed on a 
$p_z^{\rm imbalance}$-dominant segment, each MM on the atoms is very small, and the sum of the MMs is only 
one Bohr magneton. The requirement for the imbalanced states that are induced by the 
$p_z$-vacancies on one sublattice leads to the MMs of all of the $p_z^{\rm imbalance}$-dominant segments being 
in coherent FM alignment. The total MM depends on the $p_z$-vacancy concentration. These 
results explain the experimental observations of very weak magnetization with a high Curie 
temperature.

This work was supported by NFSC (No.61274097) and NBRPC (No. 2015CB921401).

\end{document}